# Magnetostructural and magnetocaloric properties of bulk LaCrO$_3$ system

Brajesh Tiwari[1#], A. Dixit[2], R Naik[3], G. Lawes[3] and M.S. Ramachandra Rao[1*]

[1]Department of Physics and Nano Functional Materials Technology Centre, Indian Institute of Technology Madras, Chennai 600 036, India

[2]Indian Institute of Technology Rajasthan, Jodhpur 341011, India

[3] Department of Physics and Astronomy, Wayne State University, Detroit, Michigan 48201, USA

#Current Address: Quantum Phenomenon and Application Division, National Physical Laboratory, New Delhi 110012, India

*Corresponding author email: msrrao@iitm.ac.in

Abstract:

We studied the magnetic properties of bulk LaCrO$_3$; a GdFeO$_3$-type distorted perovskite, with a predominant antiferromagnetic phase transition at ~ 290 K. The bulk LaCrO$_3$ exhibits intrinsic weak ferromagnetism at room temperature, which may arise due to the tilting of CrO$_6$ octahedra, resulting in a non-zero net magnetic moment, as confirmed from the magnetization measurements. A broad magnetically-induced entropy change (-ΔS) is observed with the maximum at 290 K, close to room temperature in LaCrO$_3$ system.

**Introduction:**

Transition metal oxides have been studied for many decades because of their interesting and unusual electronic and magnetic properties arising from narrow 3d bands and on-site coulomb interactions [1]. A number of these complex oxides having perovskite, or distorted perovskite structures exhibit exotic electronic and magnetic properties such as $BaTiO_3$ (ferroelectric) and $La_{1-x}Sr_xMnO_3$ (colossal magnetoresistance) with some systems, including $TbMnO_3$, and $BiFeO_3$ developing multiple ferroic orders simultaneously [2-5]. $LaCrO_3$ has been studied for several decades due to its fundamental and technological interests. Some of the earliest work in mid 1950s by Jonker et. al and Koehler et. al on structural and magnetic properties of magnetic perovskite compounds including $LaCrO_3$ [6,7]. Divalent ion substitution on La site of $LaCrO_3$ gained a lot of interest due to its high refractory nature, and a good electrical conductivity is weakly temperature dependent over a very wide range of temperatures between room temperature and 2000 K, which was found to be suitable as an electrode in magnetohydrodynamic generators [8-10] and recently as an electrode material and interconnect for fuel cells [11]. Recently Zhou et. al has studied the detailed temperature-pressure phase diagram for crystal and magnetic structures of $LaCrO_3$ by in situ neutron diffraction under pressure [12]. $LaCrO_3$ is a $GdFeO_3$–type distorted perovskite material that crystallizes in an orthorhombic crystal structure with the space group Pnma [12, 13]. The distortion in $LaCrO_3$ is due to the tilting of $CrO_6$ octahedra in opposite directions ($a^-b^+a^-$ in Glazer notation) as the Cr-O1-Cr bond angle ($160^o$) is far from the $180^o$ of an ideal perovskite. This distortion decreases the orbital overlap and width of the conduction band, leading to a non-collinear spin structure of the magnetic $Cr^{3+}$ ions resulting in weak ferromagnetism. $LaCrO_3$ exhibits antiferromagnetic ordering near room temperature [12], as confirmed in this current study. Considering the non-collinear arrangement of spins in LaCrO$_3$ with spin-orbit coupling, we observed the non-zero magnetic moment is left out in its ground-state using density functional theory [14]. An intrinsic magneto-dielectric coupling in LaCrO$_3$ has shown by temperature-dependent dielectric study in conjugation with an anomaly in optical phonon mode at antiferromagnetic ordering by Raman spectroscopy [15]. This suggests a strong spin-phonon coupling which in turns may lead to the large magnetically induced entropy change i.e. magnetocaloric effect which has magnetic refrigeration application [16]. Antiferromagnetic order, which lowers the overall symmetry cannot be created by a magnetic field from its high temperature paramagnetic phase in contrast to ferromagnetic order and the difference in symmetry between the two phases is maintained even in the presence of a magnetic-field [17]. Therefore, a higher entropy change is also expected pertaining to differences in symmetry above, and below the antiferromagnetic phase transitions and several efforts in recent times have been devoted to magnetocaloric effect study e.g. antiferromagnetic to ferromagnetic transition in RMnO$_3$ (R = Dy, Tb, Ho and Yb) compounds by Midiya et. al [18], in multiferroic BiFeO$_3$ by Ramachandran et. al [19], Ising antiferromagnet DySb by Hu et. al [20] and (Mn$_{0.83}$Fe$_{0.17}$)$_{3.25}$Ge by Du et. al [21]. In the present work, we report on structural, compositional, and magnetic properties of LaCrO$_3$ with emphasis on weak ferromagnetic properties near room temperature.

**Experimental:**

LaCrO$_3$ was synthesized using La$_2$O$_3$ (99.99 %) and Cr$_2$O$_3$ (99.9%) as starting materials by a conventional solid-state reaction method. The mixture was preheated at 600 $^o$C for 6 h prior to the calcinations. The sample was calcined twice with intermediate grinding at 850 $^o$C for about 24 h. The sample was not heated above 850 $^o$C to conserve the stoichiometry of the compound, as in this case the color of sample has been found to turn yellowish brown from the intrinsic green color, possibly due to the creation of La and O deficiencies in the lattice [22]. The powder x-ray diffraction (XRD) data of the calcined sample was collected using a PANalytical X'Pert Pro x-ray diffractometer with Cu *Kα* radiation under ambient conditions. Crystal structure refinements were carried out using General Structure Analysis System (GSAS) [23]. The elemental analysis of LaCrO$_3$ was done using a Perkin-Elmer X-ray photoelectron spectroscopy (XPS) system, equipped with cylindrical analyzer and a highly monochromatic Al *Kα* (1486.6 eV) X-ray source. The compact pellet of LaCrO$_3$ was made using hydraulic press at high pressure and mounted on sample holder using double sided carbon tape. The working pressure of the chamber was maintained at ~ $10^{-9}$ torr during the experiment. The observed binding energies of each element were identified with the standard Perkin-Elmer database [24]. The temperature and field dependent magnetic properties were measured using a physical property measurement system (PPMS, Quantum Design, USA).

**Results and discussion:**

The Rietveld refinement of LaCrO$_3$ X-ray powder diffraction data is shown in figure 1(a). The refinement was carried out using GSAS software for the orthorhombic crystal structure with the space group Pnma (# 62). The difference-profile (Diff.) between the observed (Obs.) and calculated (Calc.) diffraction pattern, as shown at the bottom of the plot. A good fit was obtained with *R* factors, w*R*p = 8.9 %, *R*p = 4.3 %, and $\chi^2$ = 1.27. The lattice constants and volume of the unit cell are found to be a = 5.479(1) Å, b = 7.759(2) Å and c = 5.516(1) Å and *V* = 234.9 Å$^3$ respectively which is in good agreement with earliar repoted values [12]. The inset in figure 1(a) shows the chemical unit cell of LaCrO$_3$ which has a total of 20 atoms (4 La, 4 Cr and 12 O) per unit cell. Each chemical unit cell of LaCrO$_3$ has corner-linked octahedra CrO$_6$ from the centers occupied by centrosymmetric Cr ions (blue) at Wyckoff position 4b (0, 0, 1/2). The corner atoms of the octahedra are oxygen ions (red) with two inequivalent positions, the apex oxygen (O1 ion) at 4c (0.498, 0.25, -0.043) and planar oxygen (O2 ion) at 8d (0.262, 0.032, -0.282). Lanthanum ions (brown) occupy the space between the octahedra at 4c (0.019, 0.25, 0.007). The distortion from ideal perovskite structure happens because of the geometric tolerance factor of 0.903 as well as antiphase tilt of adjacent octahedra which in turn lead to Cr-O1-Cr bond angle ~ 160$^o$. The XPS chemical binding energies of La, Cr, and O elements are shown in figure 1(b) and 1(c). The 3+ oxidation states of La with binding energies $3d_{5/2}$ ~ 835.2 eV and $3d_{3/2}$ ~ 851.8 eV which is due to spin-orbit coupling of La 3d states together with satellite peaks ~ 846 eV, confirms the phase purity of these $LaCrO_3$ samples [25]. A clear doublet in both La $3d_{5/2}$ and $3d_{3/2}$ is due to Coulomb energy between 3d core hole and 4f electron. The inset of figure 2(b) shows the O 1s binding energy at ~ 530.5 eV with a small feature at higher binding energies (532.5 eV) indicating the covalent nature of the bonding [26]. The $2p_{3/2}$ and $2p_{1/2}$ spin-orbit doublet components of the Cr 2p photoelectron were found at binding energies ~ 576.8 eV and ~ 586.6 eV, respectively [27]. The observed binding energies of La and Cr in $LaCrO_3$ system confirm the 3+ valence states of these elements.

Zero field cooled (ZFC), and field cooled (FC) magnetization curves of the polycrystalline $LaCrO_3$ sample were measured at a magnetic field of 1 kOe over the temperature range 10 – 380 K, as shown in of figure 2a. Curie-Weiss fit suggests a dominant antiferromagnetic ordering with very large Weiss temperature (1230 K) suggesting considerably important next neighbor interaction among Cr3+ magnetic moments. A phase transition from high temperature paramagnetic to low temperature antiferromagnetic order near $T_N$ ~ 290 K is observed as a sharp jump in magnetization. Similar but low temperature (140 K) weak ferromagnetic behavior is also observed for $YCrO_3$ and heavy rare earth chromites [28, 29]. The sharp increase in magnetization near the Néel temperature is a characteristic of weak ferromagnetism [30]. The remanent magnetization ($M_r$) and coercive field ($H_c$) of LaCrO$_3$ at different temperatures were estimated as shown in figure 2b from magnetization (M) vs applied magnetic field (H) curves measured at different temperatures (for clarity few M-H curves are shown), inset of figure 2b. Below the Néel temperature, the magnetization curves show weak ferromagnetism and there is no indication of magnetic saturation to an applied magnetic field of 50 kOe, consistent with the antiferromagnetic ordering. Abrupt jumps in the coercive field ($H_c$) and remanent magnetization ($M_r$) below the Néel temperature reflect contributions from uncompensated magnetic moments, providing support for the presence of weak ferromagnetism in LaCrO$_3$[1, 14, 15, 30].

We estimate the magnetic field-induced entropy change $\Delta S$ near the antiferromagnetic phase transition in LaCrO$_3$ using the relationship [16],

$$\Delta S(T, \Delta H) = \mu_0 \int_{H_i}^{H_f} \left(\frac{\partial M}{\partial T}\right)_H dH$$

Where $\mu_0$ is the permeability of free space. $H_i$ (considered zero in the present case) and $H_f$ are the initial and final applied magnetic fields respectively. The magnetic entropy change was estimated from the first quadrant magnetization isotherms between 310 K and 250 K in steps of 5 K (Figure 3 inset). The magnetic entropy change plotted against temperature is shown in figure 3. The maxima in isothermal magnetic entropy change $(-\Delta S)$ versus temperature T for different applied magnetic fields H, are peaked close to 290 K, which marks the onset of antiferromagnetic phase transition. It is interesting to note that upon antiferromagnetic ordering, under small applied magnetic field which don't disturb the ground state of LaCrO$_3$ shows a very small but positive entropy change which also indicates the weak ferromagnetic nature due to anisotropy in the system [31]. The maximum entropy change $(-\Delta S)$ ~ 0.11 J kg$^{-1}$K$^{-1}$ was observed at an applied magnetic field of 80 kOe. While this change is almost two orders of magnitude smaller than that observed in some other magnetocaloric materials, including Gd metal [16], understanding the mechanisms producing the MCE in LaCrO$_3$ may lead to better materials in the future. Because the magnitude of the MCE depends strongly on the sample magnetization, increasing the weak ferromagnetic moment by tuning the Cr-O1-Cr bond angle is expected to increase the change of entropy in the system. However, modifying this bond angle may also reduce the magnetic transition temperature to well below room temperature, which is detrimental to many applications. This interplay of the transition temperature and magnitude of the MCE will require careful control of bond angles in LaCrO$_3$, possibly doping with Y, to optimize the properties of the system for magnetocaloric applications.

**Conclusion:**

In conclusion, we have synthesized polycrystalline LaCrO$_3$ and confirmed an orthorhombic structure with the Pnma space group using structure refinement. Magnetization measurements reveal an antiferromagnetic transition with weak ferromagnetism at T$_N$ ~ 290 K. We observed a magnetocaloric effect near room temperature by measuring the isothermal magnetic entropy change $\Delta S(T, \Delta H)$, which shows a maximum of 0.11 J kg$^{-1}$K$^{-1}$ at a field of 80 kOe. LaCrO$_3$ a model antiferromagnetic system with weak ferromagnetism near room temperature may provide the rich physics by complex coupling among different degree of freedoms and their applications.

**Acknowledgement:**

This work was supported by the Jane and Frank Warchol Foundation together with a Career Development Chair from Wayne State University. B Tiwari and M S R Rao would to acknowledge Department of Science and Technology (DST) of India for the financial support (grant No. SR/NM/NAT-02/2005).

**List of figures**

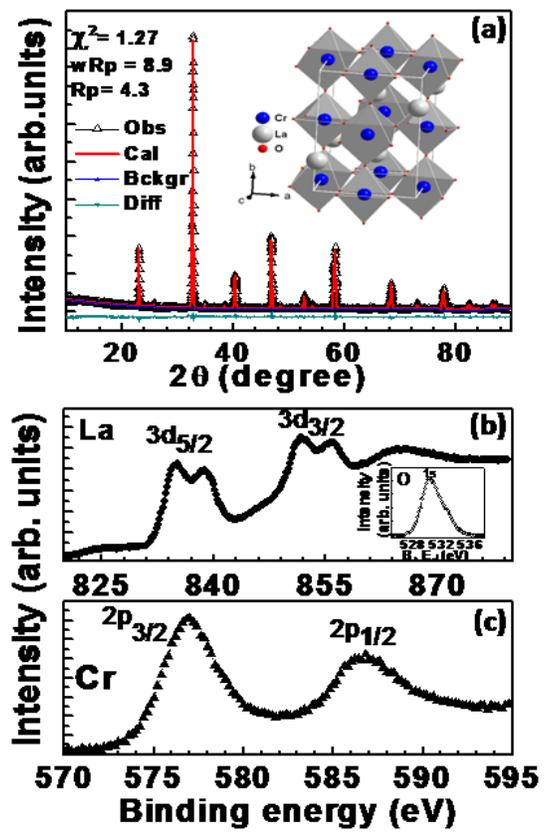

Figure 1. (a) Rietveld refined powder XRD pattern of LaCrO$_3$ sample using GSAS. Difference (Diff) between observed (Obs) and calculated (Calc) pattern is shown. Inset shows the chemical unit cell of LaCrO$_3$ where corner atoms of octahedra are oxygen ions (red), with two inequivalent positions, the apex oxygen (O1 ion) at 4c (0.498, 0.25, -0.043) and planar oxygen (O2 ion) at 8d (0.262, 0.032, -0.282) while centers of octahedra are occupied by chromium ions at 4b (0, 0, 0.5). Lanthanum ions (brown) occupy the site 4c (0.019, 0.25, 0.007), space among 8 octahedra. The X-ray photoelectron spectra as a function of binding energy (eV) of b) La 3d and O1s (inset) in LaCrO$_3$ and c) the Cr 2p doublet $2p_{3/2}$ and $2p_{1/2}$.

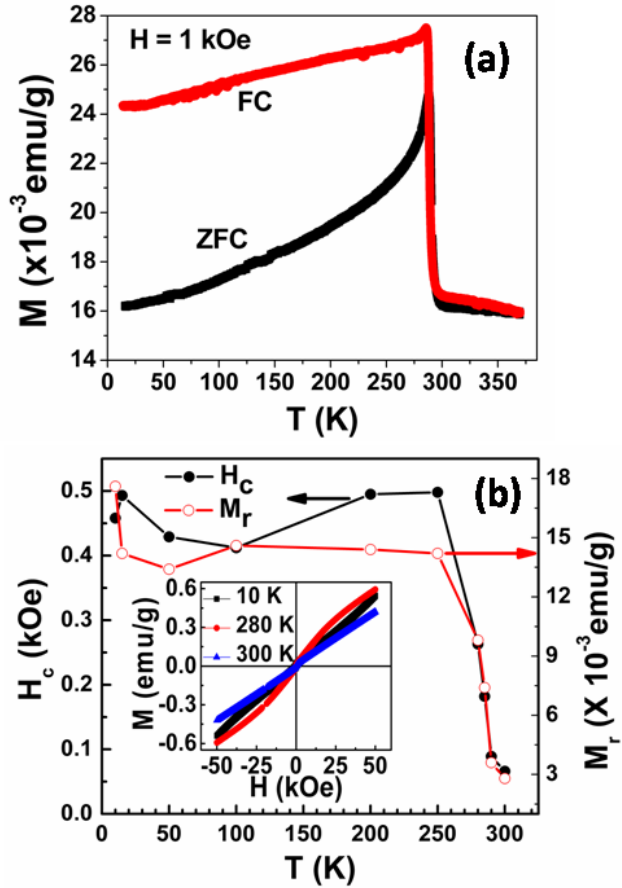

Figure 2. Temperature dependent magnetization measurements (a) zero field cooled (ZFC) and field cooled (FC) at an applied magnetic field of 1 kOe. (b) the variation of coercive field $H_c$

(solid circle, left axis) and remanent magnetization $M_r$ (open circle, right axis) of $LaCrO_3$ which were estimated from magnetization (M) vs. applied magnetic field (M-H) measurements at different temperatures, as shown inset (for clarity all M-H curves are not shown).

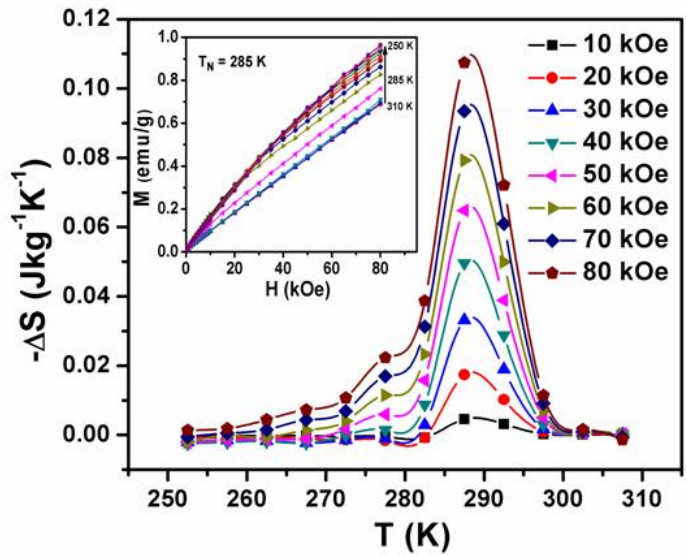

Figure 3. Temperature dependence magnetic entropy changes $(-\Delta S)$ at different magnetic fields (10 to 80 kOe) calculated from magnetization isotherms. (The solid lines are guides to the eye). Inset shows the isothermal magnetization curves at different temperatures from 250 K to 310 K in step of 5 K.